\documentclass[11pt, a4paper]{article}
\usepackage{jheppub}
\usepackage{amsfonts,amsmath,url}
\usepackage{eurosym}
\usepackage{braket}
\usepackage{graphicx}
\usepackage[utf8]{inputenc}
\usepackage{BOONDOX-cal}
\DeclareFontFamily{U}{MnSymbolC}{}
\DeclareSymbolFont{MnSyC}{U}{MnSymbolC}{m}{n}
\DeclareFontShape{U}{MnSymbolC}{m}{n}{
    <-6>  MnSymbolC5
   <6-7>  MnSymbolC6
   <7-8>  MnSymbolC7
   <8-9>  MnSymbolC8
   <9-10> MnSymbolC9
  <10-12> MnSymbolC10
  <12->   MnSymbolC12}{}
\DeclareMathSymbol{\intprod}{\mathbin}{MnSyC}{'270}

\begin{document}
\title{Corners of gravity: the case of gravity as a constrained BF theory}

\author[a]{Remigiusz Durka}
\emailAdd{remigiusz.durka@uwr.edu.pl}
\author[a,b]{Jerzy Kowalski-Glikman}
\affiliation[a]{Institute for Theoretical Physics, University of Wroc\l{}aw, pl.\ M.\ Borna 9, 50-204 Wroc\l{}aw, Poland}
\affiliation[b]{National Centre for Nuclear Research, Pasteura 7, 02-093 Warsaw, Poland}
\emailAdd{jerzy.kowalski-glikman@uwr.edu.pl}

\abstract{Following recent works on corner charges we investigate the boundary structure in the case of the theory of gravity formulated as a constrained BF theory. This allows us not only to introduce the  cosmological constant, but also explore the influence of the topological terms present in the action of this theory. Established formulas for charges resemble previously obtained ones, but we show that they are affected by the presence of the cosmological constant and topological terms. As an example we discuss the charges in the case of the AdS--Schwarzschild solution and we find that the charges give correct values.}

\keywords{boundary, gravity, corner, charges}

\maketitle
\section{Introduction}

In the traditional road to quantum gravity envisioned and formalized first by  Wheeler and De Witt \cite{Wheeler:1988zr, DeWitt:1967yk, DeWitt:1967ub, DeWitt:1967uc} the Hamiltonian constraint ${\cal H}$ is elevated to become a quantum mechanical operator $\hat{\cal H}$, which is then assumed to annihilate the physical wavefunction $\Psi$ giving rise to the celebrated Wheeler--De Witt equation 
\begin{equation}\label{WDW}
   \hat {\cal H}\Psi =0\,.
\end{equation}
This equation, along with the simpler to handle diffeomorphism and local Lorentz constraints, imposes the fundamental symmetry of gravity, the symmetry with respect to spacetime diffeomorphisms (and local Lorentz symmetry) on physical states $\Psi$. 
In the last half of a century attempts to quantize gravity concentrated on solving this equation using various techniques. In most cases, this program was implemented for spacetime having the form of a product of time and some compact spatial manifold without boundaries. In this way, the Wheeler--De Witt equation together with the equation enforcing an invariance with respect to spatial diffeomorphisms were the only conditions that the wavefunction had to satisfy, which made it possible to avoid the complications caused by the presence of the boundary data. Unfortunately, this road to quantum gravity was from the very beginning plagued by problems of both technical and conceptual nature. Technically, the Wheeler-De Witt equation is an extremely hard to solve non-linear functional differential equation with poorly understood issues of regularization and renormalization. Moreover, the meaning of solutions to this equation, `wavefunctions of the universe',  is elusive, even in the mathematically formalized framework of Loop Quantum Gravity (see \cite{Ashtekar:2021kfp} for a recent review and references therein). 

The important conceptual problem that arises for the solutions of the Wheeler--De Witt equation in the case of compact space is that this approach leaves no room for an external observer that could acquire physical information about such a universe. Indeed, since the universe has no boundary, there is nothing outside it; one can try to get information about relations between one subsystem of the universe and another, but this makes it necessary to separate the system into two subsystems, which in turn establishes a boundary between them. This simple argument makes it clear that to be able to extract physical information from the wavefunction of the universe, even in the spatially-compact case we need more information than just the fact that $\Psi$ is a solution of Wheeler--De Witt equation. 

It should be also noted in passing that the conceptual structure of the Wheeler--De Witt equation-based quantum gravity theory is very different from that of the standard perturbative quantum field theory. In the latter, following Wigner, we label solutions of equations of motion by an appropriate representation of Poincar\'e group, so that the quantum field can be interpreted as a bunch of wordlines, each carrying four-momentum and spin. Even more, you can think of a particle worldline as an interval, labeled by momentum and spin at the ends (which are to be identical as a consequence of the momentum and angular momentum conservation).  The momentum and spin are nothing but the (conserved) charges associated with the physical Poincar\'e symmetry, and following Wigner one construct Hilbert spaces as spaces on which representations of Poincar\'e group act. Inside the worldline, in its bulk, the only condition that has to be satisfied is the mass-shell condition $p^2-m^2=0$ (for external lines; for internal lines this condition is relaxed and off-shell configurations are taken into account), which, in gravity language, is essentially the Hamiltonian constraint. The mass shell condition is a constraint associated with the local diffeomorphisms of the worldline, exactly as the Hamiltonian constraint of gravity is. The end of the worldline either extends to the boundary of the system, far away from the interaction region, where the momenta and spins can be measured, or, in interacting theory, get attached to the ends of other particle wordlines. In the so formed vertex, we have to impose conservation equations. We see, therefore, that the essential difference between the bulk of the worldline and its end is that in the former only the gauge symmetry (worldline diffeomorphism) operates, while at the ends we have to deal also with physical (Poincar\'e) symmetry. 

This lesson can be extended to higher dimensional gauge  theories on spaces with boundaries. The physical difference between gauge symmetries in the bulk and physical symmetries at the boundary was stressed in \cite{Rovelli:2013fga}, \cite{Rovelli:2020mpk}, see however \cite{Riello:2021lfl} for a different view.

 In the case of (quantum) gravity, instead of compact spatial manifolds without boundary, we can consider spatial bounded regions, so that we have to do with gauge symmetries in the bulk and physical symmetries and the associated charges on the boundary. When the boundary is at spatial infinity the charges become the Poincar\'e ones, measuring the total energy, momentum and angular momentum of spacetime \cite{Regge:1974zd} or even the ones of BMS symmetry \cite{Henneaux:2018cst}. It is of interest also to consider the case when the bounded region is finite \cite{Brown:1992br}, \cite{Brown:1992bq}, \cite{Husain:1997fm}. 

In the last few years, there is a growing interest in revealing the boundary structures of gravity and their physical significance. There are several motivations for that. On the one side the discovery of the infrared triangle, a deep relation between asymptotic symmetries at null infinity (BMS symmetry \cite{Bondi:1962px}, \cite{Sachs:1962wk, Sachs:1962zza} in the case of gravity), soft theorems of quantum field theory, and memory effects (see \cite{Strominger:2017zoo} for review and references) revived the interest in charges associated with physical symmetries.

On the other side, in \cite{Freidel:2015gpa} it was shown that, quite surprisingly, the boundary symplectic structure makes the tetrad on the boundary non-commutative. It was then argued in \cite{Donnelly:2016auv} (see also \cite{Donnelly:2020xgu})  that boundary symmetries can be used to organize the Hilbert space of quantum gravity in a way similar to that Poincar\'e symmetry does in the case of quantum field theory. Then in the series of papers \cite{Freidel:2020xyx,Freidel:2020svx,Freidel:2020ayo} (see also \cite{DePaoli:2018erh} and \cite{Oliveri:2019gvm}) the boundary structure was analyzed in-depth for several gravity theories.

In this paper we investigate the boundary structure in the case of the theory of gravity formulated as a constrained BF theory \cite{Smolin:2003qu}, \cite{Freidel:2005ak}. In this formulation, having its roots in MacDowell--Mansouri construction \cite{MacDowell:1977jt}, the action of gravity is equivalent to the standard Einstein-Hilbert-Holst action with cosmological constant, appended with a number of topological invariants (see Appendix \ref{appA}). In spacetime without boundaries these topological invariants do not play any role, but when boundaries are present they influence the boundary symplectic structure and charges, making this theory potentially very different from the ones considered in \cite{Freidel:2020xyx, Freidel:2020svx, Freidel:2020ayo}. 

The formulation of gravity as a constrained BF theory \cite{Freidel:2005ak} has several advantages over the standard Plebanski formulation of Einstein-Hilbert-Holst theory \cite{Plebanski:1977zz}, in which gravity is defined as a $\sf(SO)(3,1)$ BF theory with the simplicity constraint $B = (\gamma^{-1} +\star)\, e\wedge e$ imposed, which, as we show in the appendix is effectively replaced by the constraint that makes the $B$ field proportional to (Anti) de Sitter curvature. First of all, it is manifestly local Lorentz invariant at all stages of computations of boundary symplectic structure and charges, and it does not require the troublesome imposition of simplicity constraint. Second, it contains the cosmological constant as a necessary ingredient from the very start, which makes it natural to investigate asymptotically AdS spacetimes and the associated charges \cite{Aros:1999id, Aros:1999kt, Gibbons:2004ai, Frodden:2017qwh, Frodden:2019ylc}. The charges for gravity theory given by Einstein--Cartan--Holst action appended with the topological terms were discussed in the past in \cite{Durka:2011yv, Durka:2012wd, Corichi:2016zac, Godazgar:2020gqd, Godazgar:2020kqd}.

The plan of this paper is as follows. In the following section, we present the slightly generalized constraint BF theory and show that its field equations reproduce the ones of Einstein--Cartan theory. In Section 3 we discuss the symplectic structure in the bulk of a region and on its boundary. The following section is devoted to derivation of charges associated with the Lorentz and diffeomorphism symmetries. The last section contains discussion of obtained results.

\section{Constrained BF theory with  (Anti) de Sitter gauge algebra and gravity}

In this section we present the theory of gravity formulated as a constrained BF theory with (Anti) de Sitter gauge group.

\subsection{Setup}
Our starting point is the constrained BF theory action
\begin{equation}\label{1.1}
 16 \pi \, S(A,B)= \int  F^{IJ}(A) \wedge B_{IJ}{} -
 \frac{\beta}{2} B^{IJ}\wedge B_{IJ} - \frac{\alpha}{4}{\mathcal v}_M\, \epsilon^{IJKLM} B_{IJ}
\wedge B_{KL}\, .
\end{equation}
The capital Latin letters $I,J, \ldots$ denote gauge indices of  $\sf{SO}(3,2)$ (negative cosmological constant) or $\sf{SO}(4,1)$ (positive cosmological constant) gauge algebra. The internal space vector ${\cal v}_I$ is an arbitrary unit vector, which is timelike in the flat gauge algebra metric (Killing metric) $\eta_{IJ}$, with signature $(-,+,+,+,-)$ in the case of negative cosmological constant and spacelike in the metric with signature $(-,+,+,+,+)$ in the case of positive cosmological constant\footnote{To capture both cases we introduce the signature $(-,+,+,+,\varepsilon)$.}
\begin{equation}\label{1.1a}
{\mathcal v}^I{\mathcal v}^J \eta_{IJ} =\varepsilon\,,\quad  \mbox{ with } \varepsilon = \left\{    \begin{array}{c}
-1 \mbox{ for $\sf{SO}(3,2)$} \\
1 \mbox{ for $\sf{SO}(4,1)$}
\end{array}
\right.
\end{equation}
We do not fix the non-dynamical vector ${\mathcal v}_M$ in the action \eqref{1.1}. This vector defines the Lorentz $\sf{SO}(3,1)$ subalgebra of the original (Anti) de Sitter gauge algebra as an algebra of its stability group. By allowing a generic form of this vector, constrained only by \eqref{1.1a} we are generalizing the action used in \cite{Freidel:2005ak} where the gauge fixed choice ${\mathcal v}_M=(0,\ldots,\varepsilon)$ was made.

With the help of the vector ${\cal v}_I$ we can decompose the gauge connection $A^{IJ}$ into Lorentzian $\omega^{IJ}$ and translational $e^{I}$ components as follows
\begin{equation}\label{1.1b}
	A^{IJ} = \omega^{IJ} + \frac{2\varepsilon}{\ell} e^{[I} {\mathcal v}^{J]}\,,
\end{equation}
where $\ell$ is a parameter of dimension of length needed to adjust the canonical dimension of connection (inverse length) to the one of tetrad (zero).

We further assume that the vector ${\mathcal v}^I$ is covariantly constant with respect to the $\omega$ connection \cite{Freidel:2020svx}
\begin{equation}\label{1.6}
	D^\omega {\mathcal v}^I =0
\end{equation}
and orthogonal to $e^{I}$
\begin{equation}\label{tetrad1}
e^{I} {\mathcal v}_{I}=0\,,
\end{equation}
This decomposition agrees with the one of \cite{Freidel:2005ak} for the constant vector fixed to be ${\mathcal v}^I=(0,0,0,0,1)$. Taking the $A$-covariant derivative of ${\mathcal v}^I$ we find that
\begin{equation}
    D^A{\mathcal v}^I =D^\omega {\mathcal v}^I + \frac{\varepsilon^2}{\ell}e^{I}  - \frac{\varepsilon}{\ell} e^{J} {\mathcal v}_J {\mathcal v}^{I}\,.
\end{equation}
Using \eqref{tetrad1} 
we derive
\begin{equation}\label{1.7}
	e^I =\ell\,  D^A {\mathcal v}^I\,.
\end{equation}
Clearly $e^I {\mathcal v}_{I}=\ell\,{\mathcal v}_{I} D^A {\mathcal v}^I =0$, which is consistent with our assumption \eqref{tetrad1}.

The field strength for connection $A$ is defined by standard expression
\begin{equation}\label{1.1c}
 F^{IJ} = d A^{IJ}  +  A^{I}{}_K{}\wedge A^{KJ}\,,
\end{equation}
and decomposes into
\begin{align}\label{1.1d}
   	F^{IJ} =& \left(R^{IJ} (\omega)-\frac{\varepsilon}{\ell^2}\, e^{[I}\,\wedge e^{J]}\right) + \frac{2\varepsilon}{\ell} T^{[I} {\mathcal v}^{J]}\nonumber\\
   	& +\frac{2\varepsilon}{\ell} e^{[I}\wedge D^\omega \mathcal{v}^{J]}+\frac{\varepsilon^2}{\ell^2}\left(e^{[I}\wedge e^{K]}\mathcal{v}_K\mathcal{v}^J+e^{[K}\wedge e^{J]}\mathcal{v}_K\mathcal{v}^I\right)\,.
\end{align}
The expression in the second line vanishes due to the conditions (\ref{1.6}) and (\ref{tetrad1}). In equation above we introduced the Lorentz curvature
\begin{equation}
	R^{I J}=d \omega^{I J}+\omega^{I}{}_{K}\wedge  \omega^{K J}\,,
\end{equation}
and the torsion
\begin{equation}
 T^{I}= D^{\omega} e^{I}= d  e^{I}+\omega^{I}{}_{J}\wedge  e^{J}\,.
\end{equation}
The Bianchi Identity $0= D^A F^{IJ}(A)$ decomposes into
\begin{align}\label{Bianchi_Identity}
0 &=D^{\omega}R^{IJ}\nonumber \\
&+\frac{2\varepsilon}{\ell}\left(D^{\omega
}T^{[I}-e^K\wedge R^{[I}{}_{K}\right)\mathcal{v}^{J]}\nonumber \\
&+\frac{2\varepsilon}{\ell}\left(\frac{\varepsilon}{\ell}e^{[I}\mathcal{v}^{J]}\wedge T^K \mathcal{v}_{K}-e^{[I}\wedge R^{J]K}\mathcal{v}_{K}\right)\,.
\end{align}
Two important identities can be used to simplify this expressions. First, from (\ref{1.6}) and (\ref{tetrad1}) one shows that
\begin{align}\label{TV}
T^I \mathcal{v}_I&=0\,.
\end{align}
Second, it follows from
\begin{align}
[D^\omega,D^\omega] \mathcal{v}^I=R^{IJ} \mathcal{v}_J\,.
\end{align}
that
\begin{align}\label{RV}
R^{IJ} \mathcal{v}_J&=0\,.
\end{align}
Therefore the last line of (\ref{Bianchi_Identity}) vanishes identically. Contracting now \eqref{Bianchi_Identity} with $\mathcal{v}_I$ we see that it reduces to two standard Bianchi identities
\begin{equation}\label{Bianchi_Identity1}
D^{\omega}R^{IJ}=0\,,\qquad
D^{\omega}T^{I}-R^{I}{}_{K}\wedge e^K =0\,.
\end{equation}

\subsection{Decomposition}

We make use of the vector ${\mathcal v}^I$ to decompose the $B^{IJ}$ field valued in the (Anti) de Sitter algebra into the  Lorentz and translational components (orthogonal and parallel to ${\mathcal v}^J$) {employing} the projection operator
\begin{equation}\label{1.3a}
	P^I_J\equiv \delta^I_J -\varepsilon {\mathcal v}^I{\mathcal v}_J\,,\qquad {\mathcal v}^J P^I_J={\mathcal v}_I P^I_J=0\, ,
\end{equation}
as follows
\begin{align}\label{1.3c}
	B^{IJ}_\bot & = P^I_KP^J_L\, B^{KL}=B^{IJ} -B^{IJ}_\|\,,\\
\label{1.3}
	B^{IJ}_\| & =\varepsilon\left( B^{IK}\,{\mathcal v}_K {\mathcal v}^J +B^{KJ}\,{\mathcal v}_K {\mathcal v}^I\right) \,,
\end{align}
giving rise to the decomposition
\begin{equation}\label{1.3b}
	B^{IJ}=B^{IJ}_\bot{}  +B^{IJ}_\|\, .
\end{equation}

These definitions generalize the one of \cite{Freidel:2005ak}, where ${\mathcal v}^J=(0,0,0,0,\varepsilon)$, so that the only non-zero component of $B^{IJ}_\bot{}$ are those with Lorentz indices $B^{ij}_\bot{}$ ($i,j=0,\ldots,3$), while $B^{i4}_\bot{}=0$, and the only non-zero components of $B^{IJ}_\|{}$ are those with one index equal 4, whereas $B^{ij}_\|{}=0$.

It is worth noting an alternative form 
\begin{equation}
    B^{IJ}_\bot{}=P_{K}^{I}P_{L}^{J}\,B^{KL}=-\frac{1}{4}(\epsilon ^{SQIJM}\mathcal{v}_{M})(\epsilon _{SQKLN}\mathcal{v}
^{N})B^{KL}\,.
\end{equation}
Similarly  we decompose the curvature $ F^{IJ}=F^{IJ}_\|+F^{IJ}_\bot{}$ with 
\begin{align}
	F^{IJ}_\| &=\frac{2\varepsilon}{\ell} T^{[I} \mathcal{v}^{J]}+\varepsilon (R^{IK}\,{\mathcal v}_K {\mathcal v}^J + R^{KJ}\,{\mathcal v}_K {\mathcal v}^I)\,,\\
	F^{IJ}_\bot &=\left(R^{IJ} (\omega)-\frac{\varepsilon}{\ell^2}\, e^{[I}\,\wedge e^{J]}\right) -\varepsilon (R^{IK}\,{\mathcal v}_K {\mathcal v}^J+R^{KJ}\,{\mathcal v}_K {\mathcal v}^I)\,,
\end{align}
ultimately reduced with the help of (\ref{TV}), (\ref{RV}) and a definition of the cosmological constant $\Lambda/3=\varepsilon/\ell^2$ to
\begin{align}\label{1.4a}
	F^{IJ}_\| &=\frac{2\varepsilon}{\ell} T^{[I} \mathcal{v}^{J]}\,,\\
	\label{1.4b}
	F^{IJ}_\bot &=\left(R^{IJ} (\omega)-\frac{\Lambda}{3}\, e^{[I}\,\wedge e^{J]}\right) \,.
\end{align}

\subsection{Field equations}

Let us now derive the bulk equations of motion following from the action \eqref{1.1}. Varying with respect to $B$ we find
\begin{equation}\label{1.9a}
  F^{IJ} =\beta B^{IJ} + \frac\alpha2\, {\mathcal v}_M \epsilon^{MIJ}{}_{KL}B^{KL}\, .
\end{equation}
After decomposing its inverse into the parallel and orthogonal parts we obtain
\begin{equation}\label{1.10a}
B^{IJ}_\bot +\frac{\beta^2}{\alpha^2+\beta^2}B^{IJ}_\|= \frac{\alpha}{\alpha^2+\beta^2}\,\left(\frac{\beta}{\alpha}F^{IJ}_\| + \frac{\beta}{\alpha} F^{IJ}_\bot - \frac{1}{2}\epsilon^{MIJ}{}_{KL}F^{KL}_\bot {\mathcal v}_M\right)\,,
\end{equation}
which results in
\begin{align}\label{1.10b}
	B^{IJ}_\bot &= \frac{3}{G\Lambda}\left(\gamma\,\delta^{IJ}_{KL}- \frac{1}{2}\epsilon^{MIJ}{}_{KL}{\mathcal v}_M\right)F^{KL}_\bot,\\
 B^{IJ}_\|&= \frac{3}{G\Lambda}\left(\gamma +\frac1\gamma\right)F^{IJ}_\|\label{1.10bb}\,,
\end{align}
where the projections of the curvatures are defined according to (\ref{1.4a}) and (\ref{1.4b}). The parameters $\alpha, \beta, \ell$ are related to the Newton's constant $G$, Barbero-Immirzi parameter $\gamma$ and the cosmological constant $\Lambda$ as
\begin{align}\label{constant}
   \frac{\alpha}{(\alpha^{2}+\beta^{2})}=\frac{\ell^2}{\varepsilon G}= \frac{3}{G \Lambda}\,,\qquad \frac{\beta}{\alpha}=\gamma\,,\qquad \Lambda=\varepsilon \frac{3}{\ell^2} \,,
\end{align}
and $\delta^{I J}_{K L}=\frac{1}{2}(\delta^{I}_{K}\delta^{J}_{L}-\delta^{I}_{L}\delta^{J}_{K})$. Relations \eqref{1.10b} and \eqref{1.10bb} will allow us to derive below the rest of the field equations and show in Appendix \ref{appA} the equivalence between BF (\ref{1.1}) and ECH action appended with topological terms.

Varying the action (\ref{1.1}) with respect to the connection $A$ and decomposing into the  orthogonal and parallel parts leads to
\begin{eqnarray}
0 &=&(D^{\omega }B_{\bot}^{IJ})+\frac{\varepsilon }{\ell }(e^{J}\wedge B_{\Vert}^{IK}-e^{I} \wedge B_{\Vert}^{JK}){\mathcal{v}}_{K}\,, \\
0 &=&(D^{\omega }B_{\Vert }^{IJ}){\mathcal{v}}_{J}-\frac{1}{\ell }\,e_{J}\wedge B_{\bot}^{IJ}\,,
\end{eqnarray}
which results in the equations
\begin{align}
\left(\epsilon^{IJ}{}_{KLM}T^{K}\wedge e^{L}{\mathcal{v}}%
^{M}+\frac{2}{\gamma }T^{[I}\wedge e^{J]}\right)& =0\,,\label{tor1} \\
\epsilon _{MIJKL}{\mathcal{v}}^{M}\left( R^{IJ}\wedge e^{K}-\frac{\Lambda }{3}e^{I}\wedge e^{J}\wedge e^{K} \right)+\frac{2}{\gamma }
R_{KL}\wedge e^{K}& =0\,.\label{eis1}
\end{align}
The same field equations can be obtained by substituting \eqref{1.10b} and \eqref{1.10bb} directly to the action and varying with respect to connection; see  Appendix \ref{appA}. 

It can be shown that it follows from  \eqref{tor1} that torsion vanishes
\begin{equation}\label{Tor}
T^I =0\,,
\end{equation}
and then \eqref{eis1}, by use of Biancihi identity \eqref{Bianchi_Identity1}, becomes the standard vacuum Einstein equations with cosmological constant
\begin{equation}\label{EE}
\epsilon _{MIJKL}{\mathcal{v}}^{M}\,\left(R^{IJ}\wedge e^{K}-\frac{\Lambda }{3}e^{I}\wedge e^{J}\wedge e^{K}\right) =0\,.
\end{equation}

\section{Symplectic structure in the bulk and at the corner}\label{Sect:3}

Let us now turn to the analysis of the symplectic potential of the constrained BF theory. We will make use of the covariant phase space method \cite{Kijowski:1976ze, Crnkovic:1986ex, Crnkovic:1987tz, Lee:1990nz, Iyer:1994ys, Wald:1999wa, Julia:2002df, Harlow:2019yfa}, which makes it possible to perform a canonical formalism preserving the manifest covariance at every step and which is particularly useful and straightforward in calculation of charges.

At  first sight it seems that the symplectic potential of any BF theory, constrained or not, is universally depending only on the kinetic term in the action \eqref{1.1}, but this is not really the case since the symplectic potential is computed on-shell and in this way depends on the dynamics of the theory at hands. In general
\begin{equation}\label{2.0}
\delta S = \int_M d\Theta + \ldots 
\end{equation}
where $"\ldots"$ denotes terms that vanish when the field equations are satisfied. In the case of the action \eqref{1.1} we get
\begin{equation}\label{2.1}
16 \pi \,\Theta = \int_\Sigma \delta A^{IJ}\wedge B_{IJ}\,.
\end{equation}
Using the decomposition of $A^{IJ}$ \eqref{1.1b}, equations \eqref{1.4a}, \eqref{1.4b}, \eqref{1.10b}, \eqref{1.10bb}, and taking into account that torsion vanishes, we obtain
\begin{equation}\label{2.2}
 16\pi\,	\Theta \thickapprox\frac{3}{G\Lambda}\int_{\Sigma}\delta\omega_{IJ}\wedge 
 \left(R^{KL} (\omega)-\frac{\Lambda}{3}\, e^{[K}\,\wedge e^{L]}\right) \left(\gamma \delta^{IJ}_{KL}- \frac{1}{2}\epsilon^{MIJ}{}_{KL}{\mathcal v}_M\right)\,,
\end{equation}
where $"\thickapprox"$ denotes equalities holding up to the field equations. In deriving this expression we used the identities \eqref{tetrad1} and \eqref{RV}.

Further, the term
$$
\delta \omega_{IJ} \wedge  R^{IJ} (\omega)
$$
can be expressed through the variation of the Chern--Simons Lagrangian,
 \begin{equation}\label{CS}
   \mbox{CS}(\omega) = \omega_{IJ}\wedge d \omega^{IJ} +\frac23\, \omega_{IJ}\wedge\omega^I{}_K\wedge \omega^{KJ}\,,
 \end{equation} 
 and the total derivative, to wit
 \begin{equation}\label{CS1}
\delta \omega_{IJ} \wedge  R^{IJ} (\omega)  =\frac12\, d\left(\omega_{IJ}\wedge\delta \omega^{IJ}\right) +\frac12\,\delta(\mbox{$CS(\omega)$})\,.
 \end{equation}

The term
\begin{equation}
-\frac12\,\delta \omega_{IJ} \wedge  R^{KL}\, {\mathcal v}_M\,\epsilon^{MIJ}{}_{KL}\label{3.14}
\end{equation}
is a bit more difficult to handle. Let us first consider the decomposition of its derivative part
\begin{align}
&-\frac12\delta \omega_{IJ} \wedge  d\omega^{KL}\, {\mathcal v}_M\,\epsilon^{MIJ}{}_{KL}\nonumber\\
&=\frac14\, d\left(\delta \omega_{IJ} \wedge  \omega^{KL}\, {\mathcal v}_M\,\epsilon^{MIJ}{}_{KL}  \right)+\frac14\epsilon^{MIJ}{}_{KL} \delta \omega_{IJ}  \wedge \omega^{KL}\wedge \omega_M{}^N {\mathcal v}_N\nonumber\\
  &-\frac14 \, \delta \left(\omega_{IJ} \wedge  d\omega^{KL}\, {\mathcal v}_M\,\epsilon^{MIJ}{}_{KL} \right)\,,\label{3.15}
\end{align}
where we used the fact that $D^\omega {\mathcal v}_M=0$ \eqref{1.6}. 
Notice that since ${\mathcal v}_M$ is a background field, its variation vanishes $\delta{\mathcal v}_M=0$. 

The first term on the right hand side of \eqref{3.15} is a total derivative, and it contributes to the corner symplectic potential, the second is the total variation that does not influence the symplectic form, while the two remaining ones contribute to the symplectic form on the surface $\Sigma$.

There is still a remaining non-derivative contribution to \eqref{3.14} resulting from the commutator of connections in the curvature $R^{IJ}$
$$
-\frac12\,\delta \omega_{IJ}\wedge   \omega^K{}_N\wedge\omega^{NL}{\mathcal v}_M\,\epsilon^{MIJ}{}_{KL}\,,
$$
so similarly to \eqref{CS1} we write
\begin{align}
\label{CS3}
-\frac12\,\delta \omega_{IJ} \wedge  R^{KL}\, {\mathcal v}_M\,\epsilon^{MIJ}{}_{KL}  &=-\frac{1}{4} d\left(\omega_{I J} \wedge \delta \omega^{K L}\,\mathcal{v}_{M} \epsilon^{M I J}{ }_{K L}\right)\nonumber\\
&-\frac{1}{4} \delta\left(\omega_{I J} \wedge d \omega^{K L}\,\mathcal{v}_{M} \epsilon^{M I J}{ }_{K L} \right) \nonumber \\
&-\frac{1}{2} \delta \omega_{I J} \wedge \omega^{K}{}_{N} \wedge \omega^{N L}\, \mathcal{v}_{M} \epsilon^{M I J}{}_{K L}\nonumber\\
&+\frac{1}{4}  \delta \omega_{I J} \wedge \omega^{K L} \wedge \omega_{M}{}^{N} \mathcal{v}_{N}\epsilon^{M I J}{}_{K L}\,.
\end{align}

Combining all these contributions we get the final expression for the  symplectic potential $\Theta $,  \eqref{2.2}, which decomposes into the surface $\Sigma$ contribution
\begin{align}
\Theta^\Sigma &\thickapprox - \frac{1}{16\pi G}\, \int_\Sigma  \delta \omega_{IJ}\wedge  \, e^{K}\wedge e^{L} \left(\gamma\,\delta_{KL}^{IJ}-\frac{1}{2}\,{\mathcal v}_M\,\epsilon^{MIJ}{}_{KL}  \right) \nonumber \\
  &+\frac{3}{64\pi G\Lambda}\, \int_\Sigma
  \left(\delta \omega_{IJ}  \wedge \omega^{KL}\wedge \omega_M{}^N {\mathcal v}_N 
  +2 \delta \omega_{IJ}\wedge   \omega^K{}_N\wedge\omega^{NL}{\mathcal v}_M\right)\epsilon^{MIJ}{}_{KL}\label{3.16}
\end{align}
plus an integrated  total variation and the total derivative term that produces the corner symplectic form
\begin{equation}
   \Theta^{S} \thickapprox \frac{3}{64 \pi G\Lambda}\, \int_S
  \delta\omega{}_{IJ}\wedge\omega{}^{KL}
  \left(\gamma\delta_{KL}^{IJ}-\frac{1}{2}\,{\mathcal v}_M\,\epsilon^{MIJ}{}_{KL}\right)\, . 
  \label{cornerpot}
\end{equation}

The symplectic potential on the surface $\Sigma$,  \eqref{3.16}, coincides in the case of ${\mathcal v}_M =(0,\ldots,0,\varepsilon)$  with the one derived by De Paoli and Speziale \cite{DePaoli:2018erh}, and as shown in \cite{Freidel:2020svx} is the standard gravitational symplectic potential. Indeed, for such ${\mathcal v}_M$ we have $\delta\omega_\mu{}_{i4}=e_\mu^{4}=0 $, $i=0,\ldots,3$ so the first line in \eqref{3.16} takes the standard form
\begin{equation}\label{v4symplsigma}
\frac{1}{16\pi G}\, \int_\Sigma  \delta \omega_{ij} \wedge  e^{k}\wedge e^{l} \left(\gamma\,\delta_{kl}^{ij}-\frac{1}{2}\,\epsilon^{ij}{}_{kl}  \right) \,,
\end{equation}
Moreover in this gauge the first term in the second line of \eqref{3.16} vanishes identically, while the second term combines with the last term on the right-hand side in \eqref{3.15} to become the total variation of the Chern-Simons action of connection $\tilde\omega_\mu{}^{ij} = \epsilon^{ij}{}_{kl}\, \omega_\mu{}^{kl}$ \cite{GarciaCompean:1999kj}. 

The corner $S=\partial\Sigma$ symplectic potential can be written as
\begin{equation}
   \Theta^{S} \thickapprox \frac{3}{32 \pi G\Lambda}\, \int_S  
  \left\{\delta\omega_{IJ}\wedge\omega^{KL} 
  \left(\gamma\delta_{KL}^{IJ}-\frac{1}{2}\,{\mathcal v}_M\,\epsilon^{MIJ}{}_{KL}\right) 
  \right\}\,.\label{cornerpot1}
\end{equation}
In the case of ${\mathcal v}_M =(0,\ldots,0,\varepsilon)$ the corner potential \eqref{cornerpot1} takes the form
\begin{equation}
   \Theta^{S} \thickapprox \frac{3}{32 \pi G\Lambda}\, \int_S  
  \left\{\delta\omega_{ij}\wedge\omega^{kl} 
  \left(\gamma\delta_{kl}^{ij}-\frac{1}{2}\,\epsilon^{ij}{}_{kl}\right)\right\}\label{cornerpot11}
\end{equation}
with $i,j,\ldots = 0,\ldots,3$. 

We will discuss the symplectic potential further in Section \ref{Sect:6}.

\section{Symmetries of the action}

Having derived the symplectic potentials let us now turn to the symmetries and charges associated with them. Returning to the action \eqref{1.1}
\begin{equation}\label{3.1a}
 16 \pi \, S(A,B)= \int  F^{IJ}(A) \wedge B_{IJ}{} -
 \frac{\beta}{2} B^{IJ}\wedge B_{IJ} - \frac{\alpha}{4}{\mathcal v}_M\, \epsilon^{IJKLM} B_{IJ}
\wedge B_{KL}
\end{equation}
we are going to show that it is invariant under local Anti-de Sitter $\sf{so}(3,2)$ (de Sitter $\sf{so}(4,1)$) gauge symmetry and under diffeomorphisms. Let us discuss these symmetries in detail.

If we allow the the background vector ${\mathcal v}_M$ to vary, the action  is invariant under local $\sf{so}(3,2)$ ($\sf{so}(4,1)$) transformations
\begin{align}
\delta_{\Upsilon} A^{IJ} &=D^A \Upsilon^{IJ}= \partial \Upsilon^{IJ} +A^{I}{}_K \Upsilon^{KJ}  - A^{J}{}_K \Upsilon^{KI} \nonumber\\
\delta_{\Upsilon} {\mathcal v}^I	 &=  \Upsilon^{I}{}_J\,  {\mathcal v}^J \nonumber\\
\delta_{\Upsilon} B^{IJ} &=  \Upsilon^{I}{}_K B^{KJ} - \Upsilon^{J}{}_K B^{KI} \label{symm1}
\end{align}
with {gauge parameter} $\Upsilon^{IJ}=-\Upsilon^{JI}$, where indices are lowered/raised with the help of the metric $\eta_{IJ}=(-1,1,1,1,\varepsilon)$. The first two terms of the action are obviously invariant, while the invariance of the last term follows from the fact that $\epsilon^{IJKLM}$ is an invariant tensor of
$\sf{SO}(3,2)$ ($\sf{SO}(4,1)$). 

Recalling the decomposition of $A^{IJ} = \omega^{IJ} + \frac{2\varepsilon}{\ell} e^{[I} {\mathcal v}^{J]}$ \eqref{1.1b} we can write the gauge transformation $\delta_\Upsilon A^{IJ}=D^A\Upsilon^{IJ}$ as
\begin{align}
    \delta_\Upsilon A^{IJ} &= D^\omega \Upsilon^{IJ} +\frac{2\varepsilon}{\ell} e^{[I}\mathcal{v}^{K]}\Upsilon_K{}^{J}- \frac{2\varepsilon}{\ell} e^{[K}\mathcal{v}^{J]}\Upsilon^{I}{}_{K}\,.\label{gauge1}
\end{align}
Let us now decompose the gauge parameter $\Upsilon^{IJ}$, in analogy with the connection decomposition \eqref{1.1b}, into Lorentz $\lambda^{IJ}$ and translational $\zeta^I$ parts
\begin{equation}\label{gauge2}
	\Upsilon^{IJ} =\lambda^{IJ} + 2\varepsilon \zeta^{[I}\, {\mathcal v}^{J]} \,,\qquad \lambda^{IJ} {\mathcal v}_{J}=0 = \zeta^I{\mathcal v}_I\, .
\end{equation}
Inserting this into the right hand side of \eqref{gauge1}, remembering that $D^\omega {\mathcal v}^{J}=0$, we obtain
\begin{equation}\label{gauge1a}
 	\delta_\Upsilon A^{IJ} = D^\omega \lambda^{IJ} +2\varepsilon (D^\omega\zeta^{[I}){\mathcal v}^{J]}
-\frac{2\varepsilon}{\ell}\lambda^{[I}{}_K \mathcal{v}^{J]}e^{K} +\frac{2\varepsilon}{\ell}\zeta^{[I} e^{J]}\,.
\end{equation}
As for the transformations of $\mathcal{v}^{I}$, we get from \eqref{symm1}
\begin{equation}\label{gauge1b}
 	\delta_\Upsilon \mathcal{v}^{I} = \left(\lambda^{IK} + 2\varepsilon \zeta^{[I}\, {\mathcal v}^{K]}\right) \mathcal{v}_{K} = \zeta^{I}\,.
\end{equation}
Using the connection decomposition \eqref{1.1b} and \eqref{gauge1b} we can write
\begin{equation}\label{3.1bb}
 	\delta_\Upsilon A^{IJ} = \delta_\Upsilon \omega^{IJ} + \frac{2\varepsilon}{\ell} \delta_\Upsilon e^{[I} {\mathcal v}^{J]} +\frac{2\varepsilon}{\ell} e^{[I} \zeta^{J]}\,,
\end{equation}
so that we have
\begin{equation}\label{gauge1aa}
 \delta_\Upsilon \omega^{IJ} + \frac{2\varepsilon}{\ell} \delta_\Upsilon e^{[I} {\mathcal v}^{J]}	 = D^\omega \lambda^{IJ} +2\varepsilon (D^\omega\zeta^{[I}){\mathcal v}^{J]}
-\frac{2\varepsilon}{\ell}\lambda^{[I}{}_K \mathcal{v}^{J]}e^{K} +\frac{4\varepsilon}{\ell}\zeta^{[I} e^{J]}\,.
\end{equation}
Setting $\zeta=0$ we find the standard local Lorentz transformations of Lorentz connection and tetrad
\begin{align}
	\delta_\lambda \omega^{IJ} &=D^\omega \lambda^{IJ}\label{gauge4om}\\
	\delta_\lambda e^{I} &= -\lambda^{I}{}_{K} e^K  \label{gauge4e}\, .
\end{align}
These transformations preserve the condition $D^\omega\mathcal{v}^{I}=0$. Indeed since $\delta_\lambda \mathcal{v}^{I}=0$ from \eqref{gauge2}
$$
\delta_\lambda D^\omega\mathcal{v}^{I} = \delta_\lambda \omega^I{}_J\mathcal{v}^{J} = D^\omega(\lambda^I{}_J\mathcal{v}^{J})=0\,.
$$
Let us now turn to translations generated by $\zeta$. We have
\begin{equation}\label{gaugezeta}
 \delta_\zeta \omega^{IJ} + \frac{2\varepsilon}{\ell} \delta_\zeta e^{[I} {\mathcal v}^{J]}	 = 2\varepsilon (D^\omega\zeta^{[I}){\mathcal v}^{J]}
 +\frac{4\varepsilon}{\ell}\zeta^{[I} e^{J]}
\end{equation}
with the condition 
\begin{equation}\label{gaugezetacond}
0=\delta_\zeta D^\omega\mathcal{v}^{I} =D^\omega\zeta^I +\delta_\zeta \omega^I{}_J\mathcal{v}^{J} \,.
\end{equation}
Tracing \eqref{gaugezeta} with $\mathcal{v}_{J}$, using \eqref{gaugezetacond}, $\zeta^I {\mathcal v}_{I}=0$  and $\delta e^I_\mu {\mathcal v}_{I}=-e^I_\mu\zeta_I$ following from \eqref{tetrad1}, we establish
\begin{equation}\label{gaugezetae}
   \delta_\zeta e^{I} 	 = 2\ell (D^\omega\zeta^{I})-\varepsilon e^{J}\zeta_J {\mathcal v}^{I}
\end{equation}
and then
\begin{equation}\label{gaugezetao}
   \delta_\zeta \omega^{IJ} 	 = -2\varepsilon (D^\omega\zeta^{[I}){\mathcal v}^{J]}
 +\frac{4\varepsilon}{\ell}\zeta^{[I} e^{J]}\,.
\end{equation}

Finally the $B$ field transforms as
\begin{equation}\label{gauge4B}
\delta B^{IJ} = \frac{1}{2}(\lambda^{I}{}_K B^{KJ} - \lambda^{J}{}_K B^{KI}) + \varepsilon \zeta^{[I}\mathcal{v}^{K]}B_K{}^{J}-\varepsilon \zeta^{[J}\mathcal{v}^{K]}B_K{}^{I}\, .
\end{equation}
Notice that the field strength $F^{IJ}$ transforms like the field $B$. 
\newline

Let us now turn to diffeomorphisms with infinitesimal parameter $\xi^\mu$. The transformation laws are given by the action of Lie derivative:
\begin{align}
\delta A^{IJ} &= {\cal L}_\xi A\label{diffA}\\
	\delta B_{\mu\nu}^{IJ} &= {\cal L}_\xi B^{IJ}
	\label{diffB}\\
	\delta \mathcal{v}^{I}  &= {\cal L}_\xi   \mathcal{v}^{I}\label{diffv} \, ,
\end{align}
where 
\begin{align}
    \mathcal{L}_{\xi}(\cdot)=\xi\lrcorner (d\,  \cdot)+d(\xi\lrcorner\, \cdot)\,.
\end{align}
The components of the Lorentz connection $\omega^{IJ}$ and tetrad $e^{I}$ transform under diffeomorphisms in the same way as the connection components \eqref{diffA}.

\section{Boundary charges}

Having the transformation laws, we can now compute the associated boundary charges. It would be extremely tedious to compute the charges substituting the Lorentz transformations, translations, and diffeomorphisms to the decomposed symplectic structure derived from the symplectic potentials \eqref{3.16}, \eqref{cornerpot}. Instead, we will take a step back and start with the BF-theory symplectic form
\begin{equation}\label{5.1}
16 \pi \,\Omega = \int_\Sigma \delta B_{IJ}\wedge\delta A^{IJ}\,.
\end{equation}
Having the symplectic form $\Omega$ and a symmetry $\delta_\star\,,$ the charge $\cal H$ is defined by the relation
\begin{equation}
    \delta {\cal H} = -\delta_\star \intprod \Omega\,.
\end{equation}

\subsection{Gauge symmetry charges}

For the gauge transformations \eqref{symm1} the conserved charges are 
\begin{equation}\label{5.2}
    -\delta_\Upsilon \intprod \Omega =\frac1{16 \pi } \int_\Sigma \delta B_{IJ}\wedge\delta_\Upsilon A^{IJ} - \delta_\Upsilon B_{IJ}\wedge\delta A^{IJ}\,,
\end{equation}
and are equal to the variation of the charge
$ -\delta_\Upsilon \intprod \Omega_{BF} = \delta {\cal H}[\Upsilon]$, where
\begin{equation}\label{5.3}
   {\cal H}[\Upsilon]  =\frac1{16 \pi } \int_\Sigma  B_{IJ}\wedge D^A\Upsilon^{IJ} \,.
\end{equation}
Integrating by parts, this charge can be decomposed into the bulk part
\begin{equation}\label{5.4}
   {\cal H}^\Sigma[\Upsilon]  =-\frac1{16 \pi } \int_\Sigma D^A B_{IJ} \wedge \Upsilon^{IJ} \,,
\end{equation}
which vanishes as a result of the field equations, and the corner component
\begin{equation}\label{5.5}
   {\cal H}^S[\Upsilon]  =\frac1{16 \pi } \int_S  B_{IJ}  \wedge \Upsilon^{IJ} \,.
\end{equation}
Now we can use field equations \eqref{1.10b}, \eqref{1.10bb} with \eqref{1.4a}, \eqref{1.4b} and the condition that torsion vanishes to get
\begin{equation}\label{5.6}
   {\cal H}^S[\Upsilon]  \thickapprox\frac1{16 \pi } \frac{3}{G\Lambda}\int_S \left(R^{IJ} (\omega)-\frac{\Lambda}{3}\, e^{I}\wedge e^{J}\right) \,\left(\gamma\,\delta_{IJ}^{KL} - \frac{1}{2}\epsilon_{MIJ}{}^{KL}{\mathcal v}^M\right) \Upsilon_{KL} \,.
\end{equation}
Decomposing the parameter $\Upsilon^{IJ}$  \eqref{gauge2} into translational and Lorentz components we can accordingly split the charge \eqref{5.6}. 

Using \eqref{RV} we find that the translational component of the $\sf{SO}(3,2)$ ($\sf{SO}(4,1)$) charge vanishes
\begin{equation}\label{5.8}
   {\cal H}^S[\zeta]  \thickapprox\frac\varepsilon{8 \pi } \frac{3\gamma}{G\Lambda}\int_S  R^{IJ} (\omega) \zeta_{[I}{\mathcal v}_{J]} =0\,.
\end{equation}
Therefore there is no charge associated with the translational part of the local (Anti) de Sitter symmetry. 

The  charge associated with local Lorentz symmetry has the form
\begin{equation}\label{5.7}
   {\cal H}^S[\lambda]  \thickapprox \frac{3}{16 \pi G\Lambda}\int_S  \left(R^{IJ} (\omega)-\frac{\Lambda}{3}\, e^{I}\wedge e^{J}\right) \,\left(\gamma\,\delta_{IJ}^{KL} - \frac{1}{2}\epsilon_{MIJ}{}^{KL}{\mathcal v}^M\right) \lambda_{KL} \,.
\end{equation}
We will discuss it in detail in Section \ref{Sect:6}.

\subsection{Diffeomorphism charges}

Let us now turn to the charges associated with diffeomorphisms, again following \cite{Freidel:2020xyx} and using the language of differential forms in the first few steps of derivation.
The charge associated with diffeomorphisms can be expressed as 
\begin{equation}
-\mathcal{L}_{\xi}\intprod \Omega =\frac1{16 \pi } \int_\Sigma \delta B_{IJ}\wedge\mathcal{L}_{\xi}  A^{I J}- \mathcal{L}_{\xi} B_{IJ}\wedge\delta A^{IJ}\,,
\end{equation}
which (assuming that the parameter $\xi$ is field-independent) can be rewritten
\begin{equation}
-\mathcal{L}_{\xi}\intprod \, \Omega=\frac1{16 \pi }\delta\left( \int_{\Sigma} B_{I J} \wedge \mathcal{L}_{\xi} A^{I J}\right)-\frac1{16 \pi }\int_{S} \xi\intprod\left(B_{I J} \wedge \mathcal{L}_{\xi} A^{I J}\right)\,.
\end{equation}
We assume that the only non-vanishing components of the vector field $\xi$ on the corner $S$ are those tangent to $S$. {Thus,} the second term is a diffeomorphism acting on  an integral of a scalar density and vanishes identically. Therefore the diffeomorphism charge has the form
\begin{equation}
\mathcal{H}[\xi]=\frac{1}{16 \pi }\int_{\Sigma} B_{I J} \wedge \mathcal{L}_{\xi} A^{I J}\,.\label{Hxi}
\end{equation}
Using the identity $\mathcal{L}_{\xi} A^{IJ} = \xi \lrcorner F(A)^{IJ}+D^{A}(\xi \lrcorner A^{IJ})$, which is a consequence of the Lie derivative definition $\mathcal{L}_{\xi}(\cdot)=\xi\lrcorner (d\,  \cdot)+d(\xi\lrcorner\, \cdot)$ and the field equations $D^AB_{IJ}=0$, we write \eqref{Hxi} as
\begin{equation}
\mathcal{H}[\xi]\thickapprox\frac{1}{16 \pi }\int_{\Sigma}  B_{I J} \wedge \xi \lrcorner F(A)^{IJ}+d(B_{I J} \wedge  \xi \lrcorner A^{IJ}) \,,
\end{equation}
which decomposes into the bulk part
\begin{equation}
   {\cal H}^\Sigma[\xi]  \thickapprox\frac{1}{16 \pi }\int_{\Sigma}  B_{I J} \wedge \xi \lrcorner F(A)^{IJ}
\end{equation}
and the corner component
\begin{equation}\
\mathcal{H}^{S}[\xi]=\frac{1}{16 \pi} \int_{S}  B_{I J} \wedge  \xi \lrcorner A^{IJ}\,.
\end{equation}
Using field equations we find
\begin{align}
   {\cal H}^\Sigma[\xi]  &=\frac{3}{16 \pi G\Lambda }\int_{\Sigma} \xi  \lrcorner\,\left( R^{IJ} -\frac{\Lambda}{3}\,  e^{I}\wedge e^{J}\right)\wedge \left(R_{KL} -\frac{\Lambda}{3}\, e_{K}\wedge e_{L}\right) \left(\gamma\delta_{IJ}^{KL} - \frac{1}{2}\epsilon_{MIJ}{}^{KL}{\mathcal v}^M\right).\label{diffchargeSigma}   
\end{align}
Using Bianchi identity \eqref{Bianchi_Identity1} and Einstein equations \eqref{EE} can further simplify it to give
\begin{align}
   {\cal H}^\Sigma[\xi]  &=\frac{3}{16 \pi G\Lambda }\int_{\Sigma} \left(R^{KL} -\frac{1}{2}\epsilon_{MIJKL}{\mathcal v}^M\left(R^{IJ} - 
\frac{\Lambda}{3}\, e^{I}\wedge e^{J} \right)\right)\wedge \xi \lrcorner\,R_{KL}\,.
 \label{diffchargeSigma1}   
\end{align}
 It is easy to see the charge ${\cal H}^\Sigma[\xi] $ vanishes for the vector $\xi$ tangent to the surface $\Sigma$, and therefore we do not dwell on it in this paper anymore.

As for the corner charge we obtain
\begin{equation}
\mathcal{H}^{S}[\xi]=  \frac{3}{16 \pi G\Lambda} \int_{S} \left(R^{IJ} -\frac{\Lambda}{3}\, e^{I}\wedge e^{J}\right) \,\left(\gamma\delta_{IJ}^{KL} - \frac{1}{2}\epsilon_{MIJ}{}^{KL}{\mathcal v}^M\right) \xi\lrcorner\, \omega_{KL}\,.\label{diffchargeS}
\end{equation}
This formula was previously derived in \cite{Durka:2011yv}.

\section{Discussion}\label{Sect:6}

Before we start  discussing the obtained results, let us return to the translational part of $\sf{SO}(3,2)$ ($\sf{SO}(4,1)$) gauge transformations with infinitesimal parameter $\zeta$. As we saw in Eq.\ \eqref{5.8} the charge associated with this part of gauge transformations vanishes, both on the surface $\Sigma$ and at the corner $S$. That means the translational part of gauge symmetry does not have a non-trivial physical symmetry counterpart, i.e., it is a pure gauge. For this reason, we gauge fix this symmetry by choosing the coordinates on Lie algebra of $\sf{SO}(3,2)$ ($\sf{SO}(4,1)$ ) such that ${\mathcal v}^M =(0,\ldots,0,\varepsilon)$ (at the same time setting $\epsilon^{01234}=1$ and $\epsilon_{01234}=-\varepsilon$ to write $\epsilon_{ijkl4}=\varepsilon \epsilon_{ijkl}$), which leads to the considerable simplification of the formulas in what follows.

\subsection{Symplectic structure}

As we shown in Subsection \ref{Sect:3}, the symplectic potentials on the surface $\Sigma$ and at the corner $S$ take the form
\begin{equation}\label{v4symplsigma1}
\Theta^{\Sigma} \thickapprox \frac{1}{16\pi G}\, \int_\Sigma  \delta \omega_{ij} \wedge  e^{k}\wedge e^{l} \left(\gamma\,\delta_{kl}^{ij}-\frac{1}{2}\,\epsilon^{ij}{}_{kl}  \right) \,,
\end{equation}
and
\begin{equation}
   \Theta^{S} \thickapprox \frac{3}{32 \pi G\Lambda}\, \int_S  
  \delta\omega_{ij}\wedge\omega^{kl} 
  \left(\gamma\delta_{kl}^{ij}-\frac{1}{2}\,\epsilon^{ij}{}_{kl}\right)\, ,
  \label{cornerpot111}
\end{equation}
where we use the gauge indicies $i,j,\ldots = 0,\ldots,3$. The expression \eqref{v4symplsigma1} agrees with the standard symplectic potential of Einstein--Cartan--Holst theory \cite{Freidel:2020svx}, while the corner component of the latter, contrary to what we have here \eqref{cornerpot111}, is zero. Let us now discuss these formulas.

It can be shown that the first term in \eqref{v4symplsigma1} gives only the corner contribution. Indeed since torsion vanishes, we have
\begin{align}
0&=    \delta T^i \wedge e_i = d\delta e^i \wedge e_i + \delta\omega^i{}_j\wedge e^j \wedge e_i+ \omega^i{}_j\wedge \delta e^j \wedge e_i\nonumber\\
&= d\left(\delta e^i \wedge e_i\right) - \delta e^i \wedge\omega_{ij}\wedge e^j+ \delta\omega^i{}_j\wedge e^j \wedge e_i+ \omega^i{}_j\wedge \delta e^j \wedge e_i\nonumber\\
&= d\left(\delta e^i \wedge e_i\right) + \delta\omega^i{}_j\wedge e^j \wedge e_i\,,\nonumber
\end{align} 
so that we get
\begin{equation}\label{v4symplsigma2}
\Theta^{\Sigma} \thickapprox-\frac{1}{32\pi G}\, \int_\Sigma  \delta \omega_{ij} \wedge  e^{k}\wedge e^{l}\, \epsilon^{ij}{}_{kl}   \,,
\end{equation}
and
\begin{equation}
   \Theta^{S} \thickapprox \frac{3}{32 \pi G\Lambda}\, \int_S  
  \delta\omega_{ij}\wedge\omega^{kl} 
  \left(\gamma\delta_{kl}^{ij}-\frac{1}{2}\,\epsilon^{ij}{}_{kl}\right) + \frac{\gamma}{16\pi G}\, \int_S  \delta e^i \wedge e_i\, .
  \label{cornerpot112}
\end{equation}

In the case of Einstein--Cartan--Holst theory discussed in \cite{Freidel:2020xyx,Freidel:2020svx,Freidel:2020ayo} the starting point to compute the symplectic potential is, as it was in our case, the universal formula \eqref{2.1} (with (Anti) de Sitter indices $I,J$ replaced by the Lorentz ones $i,j$) to which one inserts the simplicity constraint
$$
B_{ECH}^{ij}=\frac1{16\pi G} \left(\gamma^{-1}\,\delta_{kl}^{ij}-\frac{1}{2}\,\epsilon^{ij}{}_{kl}  \right)e^{i}\wedge e^{j}\,.
$$
Therefore in the ECH case the symplectic potential over the surface $\Sigma$ agrees with the one of the present paper, while the corner ones are different. This comes as no real surprise, since the topological terms present in constrained BF theory action influence the form of the symplectic potential.

\subsection{Charges}

Let us start rewriting the charges \eqref{5.7}, \eqref{diffchargeSigma}, \eqref{diffchargeS} using previously mentioned gauge fix ${\mathcal v}^I =(0,0,0,0,\varepsilon)$. For the Lorentz charge at the corner we have
\begin{equation}\label{LorentzS}
   {\cal H}^S[\lambda]  \thickapprox \frac{3}{16 \pi G\Lambda}\int_S  F^{ij} \,\left(\gamma\,\delta_{ij}^{kl} - \frac{1}{2}\epsilon_{ij}{}^{kl}\right) \lambda_{kl} \,,
\end{equation}
while for the diffemorphism charge we obtain
\begin{align}
   {\cal H}^\Sigma[\xi]  &=\frac{3}{16 \pi G\Lambda }\int_{\Sigma} \xi  \lrcorner\, F^{ij} \wedge F_{kl}  \left(\gamma\,\delta_{ij}^{kl} - \frac{1}{2}\epsilon_{ij}{}^{kl}\right)\label{diffchargeSigma2}   
\end{align}
for the surface $\Sigma$, and 
\begin{equation}\
\mathcal{H}^{S}[\xi]=  \frac{3}{16 \pi G\Lambda} \int_{S} F^{ij}  \,\xi\lrcorner\, \omega_{kl}\left(\gamma\,\delta_{ij}^{kl} - \frac{1}{2}\epsilon_{ij}{}^{kl}\right)  \label{diffchargeS1}
\end{equation}
for the corner $S$, 
where for simplicity we use the (Anti) de Sitter curvature 
\begin{equation}
    F_{ij}= R_{ij} -\frac{\Lambda}{3}\, e_{i}\wedge e_{j}\,.\label{AdS-curvature}
\end{equation}

Thanks to a particular form of the starting action, we can use the identity $$(\xi  \lrcorner\, F^{ij}) \wedge F_{kl}(\gamma\,\delta_{ij}^{kl} - \frac{1}{2}\epsilon_{ij}{}^{kl}) = 1/2\,\xi  \lrcorner\,\left( F^{ij} \wedge F_{kl}  (\gamma\,\delta_{ij}^{kl} - \frac{1}{2}\epsilon_{ij}{}^{kl}) \right)$$ to rewrite the surface $\Sigma$ diffeomorphism charge simply as
\begin{align}
   {\cal H}^\Sigma[\xi]  &=\frac{1}{2 }\int_{\Sigma} \xi  \lrcorner\, L\,, \label{diffchargeSigma3}   
\end{align}
where $L$ is the Lagrangian of the theory
\begin{equation}
    L = \frac{3}{16 \pi G\Lambda }\, F^{ij} \wedge F_{kl}  \left(\gamma\,\delta_{ij}^{kl} - \frac{1}{2}\epsilon_{ij}{}^{kl}\right)\,.
\end{equation}
As we noticed above the charge ${\cal H}^\Sigma[\xi]$ vanishes for $\xi$ tangential to $\Sigma$.

There are three immediate properties of the corner charges ${\cal H}^S[\lambda]$, ${\cal H}^S[\xi]$ that are worth noting. First, these charges vanish identically in a vacuum, i.e.\ for the (Anti) de Sitter space that satisfies the equation $F_{ij}=0$. This is a very desirable feature, because we expect that all the charges of the vacuum should be zero. Contrary to that the charges of Einstein-Cartan-Holst theory can be non-zero even in the case of flat spaces with vanishing curvature.

Second, the corner charges \eqref{LorentzS} and \eqref{diffchargeS1}  are essentially identical in form, in a sense that for a given diffeomorphism $\xi$ tangential to a corner $S$ and a given configuration of Lorentz connection $\omega^{ij}$ there always exists a Lorentz symmetry parameter $\lambda^{ij}$ giving the same corner charge value.

Last but not least, we should observe that for the charges \eqref{LorentzS}--\eqref{diffchargeS1} there is no smooth flat space limit $\Lambda \rightarrow0$. At first sight, this may seem surprising because for the field equations \eqref{EE} a smooth vanishing cosmological constant limit exists. The reason why this is not the case for the charges is that for the topological terms that are part of the action \eqref{actionA.11} such limit does not exist. These topological terms do not affect the field equations, but they contribute to the boundary terms. This shows that gravity defined as a constrained BF theory is fundamentally different from the theory in Einstein--Cartan--Holst formulation.

We can compute the algebra of charges using the fact {that by construction} the charges generate infinitely many transformations through the Poisson bracket
\begin{equation}
    \left\{ {\cal H}^S[\Xi], (\star)\right\} = \delta_\Xi (\star)\,.
\end{equation}
The Poisson bracket of two charges can be expressed as variation of the charge
\begin{equation}
    \left\{ {\cal H}^S[\Xi_1], {\cal H}^S[\Xi_2]\right\} =\frac12\left( \delta_{\Xi_1}  {\cal H}^S[\Xi_2]
-\delta_{\Xi_2}  {\cal H}^S[\Xi_1]\right)\,.
\end{equation}
For two Lorentz charges we have
\begin{align}
 \left\{ {\cal H}^S[\lambda_1], {\cal H}^S[\lambda_2]\right\}  &=   \frac{3}{16 \pi G\Lambda}\int_S  \left(\lambda_1{}^i{}_mF^{mj} + \lambda_1{}^j{}_mF^{im} \right)\left(\gamma\,\delta_{ij}^{kl} - \frac{1}{2}\epsilon_{ij}{}^{kl}\right) \lambda_2{}_{kl} \nonumber\\
 &= \frac{3}{16 \pi G\Lambda}\int_S  F^{ij} \left(\gamma\,\delta_{ij}^{kl} - \frac{1}{2}\epsilon_{ij}{}^{kl}\right) \left(\lambda_1{}_k{}^m\lambda_2{}_{ml} + \lambda_1{}_l{}^m\lambda_2{}_{km} \right)\nonumber\\
 &= {\cal H}^S\left[\,[\lambda_1, \lambda_2]\,\right]\,,
\end{align}
so that Poisson bracket of two Lorentz charges is a Lorentz charge associated with the commutator of two original Lorentz symmetry parameters.

Analogously we check the the bracket of two diffeomorphism charges is the diffeomorphism charge computed with respect to the vector field $\xi = [\xi_1,\xi_2]$
\begin{align}
 \left\{ {\cal H}^S[\xi_1], {\cal H}^S[\xi_2]\right\}  &=   {\cal H}^S\left[\,[\xi_1, \xi_2]\,\right]
\end{align}
and that for mixed commutator
\begin{align}
 \left\{ {\cal H}^S[\xi], {\cal H}^S[\lambda]\right\}  &=   {\cal H}^S\left[\xi\lrcorner\,d\lambda \right]\,.
\end{align}
That is the  $\mbox{diff}(S) \ltimes \sf{SO}(3,1)$ corner algebra as in the case of Einstein-Cartan-Holst theory.

In the discussion below we will consider the corner charges associated with finite regions, as well as the asymptotic ones. Since the asymptotic structure in the case of negative cosmological constant (Anti-de Sitter space) and the positive one (de Sitter space) are very different, we will consider here only the former (see \cite{Kolanowski:2021hwo} for  recent discussion of charges in asymptotically de Sitter space.)

\section{Example: charges of AdS--Schwarzschild}

We complete this paper showing some explicit expressions for the charges, whose general form was analyzed in the preceding section. To this end we consider the AdS--Schwarzschild spacetime (the charges of AdS-Kerr and AdS-Taub-NUT have similar structure and we will discuss them in a separate paper), which is is defined by the tetrad one forms
\begin{align}
  e^0=f(r) dt,\qquad e^1=f(r)^{-1}dr,\qquad e^2=r d\theta, \qquad e^3=r \sin\theta d\varphi
\end{align}
with
\begin{equation}
    f(r)^{2}=\frac{r^{2}-2 G M r- r^{4}\Lambda/3}{r^{2}}\,.
\end{equation}
The non-zero spin connection components are given by
\begin{align}
\omega^{01} &=f'(r) e^{0} &\omega^{23} &=-\frac{1}{r} \frac{\cos\theta}{\sin\theta} e^{3}\\
\omega^{12} &=-\frac{f(r)}{r} e^{2} &\omega^{13} &=-\frac{f(r)}{r} e^{3}\,,
\end{align}
whereas AdS-curvatures \eqref{AdS-curvature} can be expressed in terms of  vierbeins as follows
\begin{align}\label{AdS-spanned}
F^{01}&=-2 h e^{0} \wedge e^{1} &F^{02}&=h e^{0} \wedge e^{2} \nonumber\\
F^{03}&=h e^{0} \wedge e^{3} &F^{12}&=h e^{1} \wedge e^{2}\nonumber\\
F^{13}&=h e^{1} \wedge e^{3} &F^{23}&=-2 h e^{2} \wedge e^{3}
\end{align}
with the auxiliary function $h$ taking a simple form $h=-\frac{G M}{r^3}$.

Let us first consider the corner charge of Lorentz symmetry \eqref{LorentzS}. In the coordinates $(t,r,\theta,\varphi)$ in which the AdS--Schwarzschild solution is written down, the corner $S$ is defined by conditions $t=$const, $r=$const. One finds that the Lorentz charge is a combination of a radial boost  and rotation of the sphere $S$
\begin{align}\label{LorentzS1}
   {\cal H}^S[\lambda]  &\thickapprox \frac{3}{4 \pi G\Lambda}\int_S   h \left(\lambda_{01} -\gamma  \lambda_{23}\right)e^2\wedge e^3\nonumber\\
      &\thickapprox \frac{3}{4 \pi G\Lambda}\int_S  \frac{-G M}{r^3} \left( \lambda_{01} -\gamma  \lambda_{23}\right)  r^2 \sin\theta d\theta \wedge d\varphi\,.
\end{align}
It is easy to see that asymptotically, when $r=\infty$, all the Lorentz charges vanish, whereas for finite $r= R$:
\begin{align}
   {\cal H}^S[\lambda]  &\thickapprox- \frac{1}{4 \pi} \frac{3}{\Lambda}\frac{M}{R} \int_S   \left(\lambda_{01} -\gamma  \lambda_{23}\right)  \sin\theta d\theta \wedge d\varphi \,.
\end{align}
Comparing \eqref{diffchargeS1} with \eqref{LorentzS} we see that for the corner diffeomorphism charges we have 
\begin{align}\label{LorentzS2}
   {\cal H}^S[\xi]  &= \frac{3}{4 \pi G\Lambda}\int_S  h \left(\omega^\xi_{01}- \gamma \omega^\xi_{23}\right) e^2\wedge e^3  \,,
\end{align}
where
\begin{equation}
    \omega^\xi_{ij} = \xi \lrcorner\,\omega_{ij}
\end{equation}
with $\xi$ being a unit vector tangent to the corner $S$
\begin{equation}
    \xi = \xi^\theta\, \frac{1}{r} \frac{\partial}{\partial\theta} + \xi^\phi\, \frac{1}{r \sin\theta} \frac{\partial}{\partial\phi}\,,\quad (\xi^\theta)^2 +(\xi^\phi)^2=1\,.
\end{equation}
Notice that in case of such $\xi$ for the sphere $S$ at $r=R$ we find
\begin{align}
   {\cal H}^S[\xi]  &= \frac{3\gamma}{4 \pi G\Lambda}\int_S  (G M/R^3 )\, \omega^\xi_{23} \, R^2 \sin\theta d\theta \wedge d\varphi\nonumber\\
     &= -\frac{3\gamma}{4 \pi\Lambda}\frac{M}{R^2}\int_S     \xi^\phi \cos\theta\, d\theta \wedge d\varphi\,.
\end{align}
Decomposing $\xi^\phi$ into spherical harmonics, we see that the integral is non-vanishing only for $\xi^\phi =4\pi Y^{10} \sim \cos\theta$.

Expressions above contain the inverse of the cosmological constant, which could be traced down to the coupling constant in front of the topological Euler term. Let us point out that the topological terms are fixed within  the constrained BF theory. It turns out that the boundary contributions resulting from the topological terms  regularize the charges, which are interpreted as as mass or angular momentum of spacetime \cite{Aros:1999id, Aros:1999kt, Durka:2011yv, Durka:2012wd}. For instance for AdS--Schwarzschild spacetime and the action rewritten in the symbolic form
\begin{align}
32 \pi G S=\int \text{ECH}-\frac{\Lambda}{6} \int \text{ cosmological }+\rho \left(\int\text { Euler }+ 2\gamma \int\text { Pontryagin }\right)\,,
\end{align}
it can be shown that unless we fix $\rho$ to have exactly the value of MacDowell-Mansouri model \cite{MacDowell:1977jt} related to the cosmological constant, we would face the problems of wrong coefficient multiplying the mass and the expression that is divergent
\begin{align}
    \text { Mass }=Q\left(\partial_{t}\right)=\frac{M}{2}\left(1-\frac{2 \Lambda}{3} \rho\right)-\lim _{r \rightarrow \infty} \frac{\Lambda}{3}\frac{r^{3}}{2G}\left(1+\frac{2 \Lambda}{3} \rho\right)\,.
\end{align}
 In our BF theorym with $\rho=-\frac{3}{2\Lambda}$, we are not only avoiding such divergence, but we also provide correct the regularization of the values of charges through the topological terms, at the same time assuring additional complementary Holst--Pontryagin regularization (see for more details \cite{Aros:1999id, Aros:1999kt, Durka:2011yv, Durka:2012wd}). For completeness, if we are interested in the time-like vector $\xi^t=1$ alone, the formula \eqref{diffchargeS} at infinity gives the mass
\begin{align}
 \mathcal{H}^{S}[\xi^t] =\frac{1}{4\pi G}\lim_{r\to \infty}\int_S  \left(\frac{G M}{r^2}-\frac{3(G M)^2}{\Lambda r^5} \right) \,e^2 \wedge e^3=M\,.
\end{align}
At the same time for the AdS--Schwarzschild spacetime a rotational part calculated for the vector $\xi^\varphi=1$ vanishes. 

Of interest is also the charge associated with the boundary  located at the horizon $H$ defined to be given by solution of the equation $r_{+}-2G M - \frac{\Lambda}{3} r_{+}^3=0$
\begin{align}
   {\cal H}^H[\lambda]  &\thickapprox \frac{1}{4 \pi} \frac{3}{\Lambda}\frac{M}{r_{+}} \int_S   \left(\gamma  \lambda_{23}-\lambda_{01}\right)  \sin\theta d\theta \wedge d\varphi \,.
\end{align}

We leave for the future work the analysis of  AdS--Kerr and  AdS--Taub-NUT spacetimes \cite{Misner:1963fr, Durka:2019ajz}, where we expect more interesting outcomes.

\section*{Acknowledgment}
We would like to thank Laurent Freidel and Daniele Pranzetti for discussions and comments on the early version of this manuscript. 

For JKG, this work was supported by funds provided by the National Science Center, projects number 2017/27/B/ST2/01902 and 2019/33/B/ST2/00050.

\appendix
\section{From constrained BF to Einstein-Cartan-Holst action}\label{appA}

The action of gravity described by the constrained BF theory is a combination of the Cartan–Einstein action with a cosmological constant term and the Holst term, accompanied by the topological Euler, Pontryagin and Nieh-Yan terms. To show this explicitly, one starts with the action  \eqref{1.1} and solves it for $B^{IJ}$. Using  $B$ field equation
\begin{align}\label{constraint}
      F^{IJ} =\beta B^{IJ} + \frac\alpha2\, {\mathcal v}_M \epsilon^{MIJ}{}_{KL}B^{KL} 
\end{align}
and plugging it to the second and third term in \eqref{1.1}, a new form of the action is just a half of the $BF$ action
\begin{equation}
16 \pi\,S(A,B)   =\frac{1}{2}\int B^{IJ}\wedge F_{IJ}\,,\label{halfBF}
\end{equation}
accompanied by the constraint relating $B^{IJ}$ with $F^{IJ}$  \eqref{constraint}. Let us notice at this point that in the case of Einstein--Cartan--Holst theory discussed in \cite{Freidel:2020xyx,Freidel:2020svx,Freidel:2020ayo} we have to do with the same  action \eqref{halfBF}, but instead of \eqref{constraint} one takes the Lorentzian curvature $F_{ij}=R_{ij}(\omega)$ and employs the simplicity constraint 
\begin{equation}\label{simpconstraint}
    B^{ij} =\frac{1}{16\pi G} \left(\gamma^{-1}\delta^{ij}_{kl}+ \frac{1}{2}\epsilon^{ij}{}_{kl}\right) e^i\wedge e^j\,
\end{equation}
is employed.

Now it is just a matter of substitution of the field equations for components of the field $B^{IJ}=B^{IJ}_\bot+ B^{IJ}_\|$,
\begin{align}
B^{IJ}_\bot&= \frac{\alpha}{(\alpha^{2}+\beta^{2})}\left(\frac{\beta}{\alpha}\,\delta^{IJ}_{KL}- \frac{1}{2}\epsilon^{MIJ}{}_{KL}{\mathcal v}_M\right)F^{KL}_\bot\\
B^{IJ}_\| &= \frac{\alpha}{(\alpha^{2}+\beta^{2})}\left(\frac{\beta}{\alpha} +\frac{\alpha}{\beta}\right)F^{IJ}_\|\,,
\end{align}
to obtain
\begin{align}
16 \pi\,S(A) &  =\frac{1}{2}\frac{\alpha}{(\alpha^{2}+\beta^{2})}\int
	\left(\frac{\beta}{\alpha} +\frac{\alpha}{\beta}\right)F^{IJ}_\| \wedge ( F_{\bot~IJ}+  F_{\|~IJ}) \nonumber\\
	&+\left(\frac{\beta}{\alpha}\,\delta^{IJ}_{KL}- \frac{1}{2}\epsilon^{MIJ}{}_{KL}{\mathcal v}_M\right)F^{KL}_\bot \wedge ( F_{\bot~IJ}+  F_{\|~IJ})\,,
\end{align}
where particular components of curvature are expressed as $F_{\bot }^{IJ}=R^{IJ}(\omega )-\frac{\varepsilon }{\ell ^{2}}\,e^{I}\wedge e^{J}$  and $F_{\Vert }^{IJ}=\frac{2\varepsilon }{\ell }T^{[I}
\mathcal{v}^{J]}$. Using orthogonality
\begin{align}
F_{\Vert}^{IJ}\wedge F_{\bot~IJ} & = \frac{2\varepsilon}{\ell}T^{I}\wedge R_{IJ}\mathcal{v}^{J}=0\,\\
F_{\Vert}^{KL}\wedge F_{\bot~IJ}& \,\epsilon^{MIJ}{}_{KL}{\mathcal v}_M=0\,,
\end{align}
leads to a concise form that could seen as the generalized MacDowell-Mansouri construction \cite{MacDowell:1977jt}
\begin{align}
16\pi\,S(A) &  =\frac{1}{2}\frac{\alpha}{(\alpha^{2}+\beta^{2})}\int
	((\frac{\alpha}{\beta}+\frac{\beta}{\alpha})F_{\Vert}^{IJ}\wedge F_{\Vert~IJ}\nonumber\\
	&+\left(\frac{\beta}{\alpha}\delta^{IJ}_{KL}-\frac{1}{2}{\epsilon^{IJ}{}_{KLM}\mathcal{v}}^{M})F_{\bot}^{KL}\wedge F_{\bot~IJ}\right)\,,
\end{align}
Substituting
\begin{align}
F_{\Vert}^{IJ}\wedge F_{\Vert~IJ} & =\frac{2\varepsilon^{3}}{\ell^{2}}T^{I}\wedge T_{I}\,\\
F_{\bot}^{IJ}\wedge F_{\bot~IJ} & = R^{IJ}\wedge R_{IJ}-\frac{2\varepsilon}{\ell^2}R_{IJ}\wedge e^{[I}\wedge e^{J]}\,,
\end{align}
and using explicit expressions for $\alpha ,\beta, \ell $ constants \eqref{constant} we end up with
\begin{align}
32\pi G\,S(A)& =\int {\mathcal{v}}^{M}\epsilon
{}_{MIJKL}\left(R^{IJ}\wedge e^{K}\wedge \,e^{L}-\frac{\Lambda}{6} e^{I}\,\wedge e^{J}\wedge e^{K}\,\wedge
e^{L}\right) \nonumber \\
& +\frac{2}{\gamma }R_{IJ}\wedge e^{I}\wedge e^{J} \nonumber \\
& +2 \left(\frac{1}{\gamma }+\gamma\right)(T^{I}\wedge T_{I}-R_{~IJ}\wedge e^{I}\wedge
e^{J}) \nonumber \\
& +\gamma \frac{3}{\Lambda}R^{IJ}\wedge R_{IJ}-\frac{3}{2\Lambda}{\mathcal{v}}^{M}\epsilon {}_{MIJKL}R^{IJ}\wedge R^{KL}\,.\label{actionA.11}
\end{align}
We see that the    constrained $BF$ model
is described by the action containing the Cartan--Einstein action with the cosmological
constant term,  the Holst term and a combination of the 
Nieh-Yan, Pontryagin, and Euler topological terms. The field equations yields exactly of form \eqref{tor1} and \eqref{eis1}. The vanishing of torsion following from the first equation makes the Holst contribution vanish as a consequence of Bianchi identity. Note that we do not have contributions to the field equations from the rest of the terms. The Nieh-Yan, and Pontryagin can be expressed as the total derivatives 
\begin{align}
NY& =\int d\left( e_{I}\wedge T^{I}\right) =\int d\left( e_{I}\wedge
D^{\omega }e^{I}\right)  \\
P& =\int R_{IJ}\wedge R^{IJ}=\int d(C(\omega )\,),
\end{align}%
where $C(\omega)$ is expressed by the Chern-Simons term \eqref{CS}. As for the Euler term, let us just mention that its variation vanishes due to the Bianchi identity.


\begin{thebibliography}{99.}

\bibitem{Wheeler:1988zr}
J.~A.~Wheeler,
``Superspace and the nature of quantum
geometrodynamics", in C DeWitt and JW  Wheeler, editors, Batelle
Rencontres: 1967 Lectures in Mathematics and Physics, 242.  Benjamin,
New York, 1968; reprinted in 
Adv. Ser. Astrophys. Cosmol. \textbf{3} (1987), 27-92

\bibitem{DeWitt:1967yk}
B.~S.~DeWitt,
``Quantum Theory of Gravity. 1. The Canonical Theory,''
Phys. Rev. \textbf{160} (1967), 1113-1148
doi:10.1103/PhysRev.160.1113

\bibitem{DeWitt:1967ub}
B.~S.~DeWitt,
``Quantum Theory of Gravity. 2. The Manifestly Covariant Theory,''
Phys. Rev. \textbf{162} (1967), 1195-1239
doi:10.1103/PhysRev.162.1195

\bibitem{DeWitt:1967uc}
B.~S.~DeWitt,
``Quantum Theory of Gravity. 3. Applications of the Covariant Theory,''
Phys. Rev. \textbf{162} (1967), 1239-1256
doi:10.1103/PhysRev.162.1239

\bibitem{Ashtekar:2021kfp}
A.~Ashtekar and E.~Bianchi,
``A short review of loop quantum gravity,''
Rept. Prog. Phys. \textbf{84} (2021) no.4, 042001
doi:10.1088/1361-6633/abed91

\bibitem{Rovelli:2013fga}
C.~Rovelli,
``Why Gauge?,''
Found. Phys. \textbf{44} (2014) no.1, 91-104
doi:10.1007/s10701-013-9768-7
[arXiv:1308.5599 [hep-th]].

\bibitem{Rovelli:2020mpk}
C.~Rovelli,
``Gauge Is More Than Mathematical Redundancy,''
Fundam. Theor. Phys. \textbf{199} (2020), 107-110
doi:10.1007/978-3-030-51197-5\_4
[arXiv:2009.10362 [hep-th]]

\bibitem{Riello:2021lfl}
A.~Riello,
``Edge modes without edge modes,''
[arXiv:2104.10182 [hep-th]].

\bibitem{Regge:1974zd}
T.~Regge and C.~Teitelboim,
``Role of Surface Integrals in the Hamiltonian Formulation of General Relativity,''
Annals Phys. \textbf{88} (1974), 286
doi:10.1016/0003-4916(74)90404-7

\bibitem{Henneaux:2018cst}
M.~Henneaux and C.~Troessaert,
``BMS Group at Spatial Infinity: the Hamiltonian (ADM) approach,''
JHEP \textbf{03} (2018), 147
doi:10.1007/JHEP03(2018)147
[arXiv:1801.03718 [gr-qc]].

\bibitem{Brown:1992br}
J.~D.~Brown and J.~W.~York, Jr.,
``Quasilocal energy and conserved charges derived from the gravitational action,''
Phys. Rev. D \textbf{47} (1993), 1407-1419
doi:10.1103/PhysRevD.47.1407
[arXiv:gr-qc/9209012 [gr-qc]].

\bibitem{Brown:1992bq}
J.~D.~Brown and J.~W.~York, Jr.,
``The Microcanonical functional integral. 1. The Gravitational field,''
Phys. Rev. D \textbf{47} (1993), 1420-1431
doi:10.1103/PhysRevD.47.1420
[arXiv:gr-qc/9209014 [gr-qc]].

\bibitem{Husain:1997fm}
V.~Husain and S.~Major,
``Gravity and BF theory defined in bounded regions,''
Nucl. Phys. B \textbf{500} (1997), 381-401
doi:10.1016/S0550-3213(97)00371-4
[arXiv:gr-qc/9703043 [gr-qc]].

\bibitem{Bondi:1962px}
H.~Bondi, M.~G.~J.~van der Burg and A.~W.~K.~Metzner,
``Gravitational waves in general relativity. 7. Waves from axisymmetric isolated systems,''
Proc. Roy. Soc. Lond. A \textbf{269} (1962), 21-52
doi:10.1098/rspa.1962.0161

\bibitem{Sachs:1962wk}
R.~K.~Sachs,
``Gravitational waves in general relativity. 8. Waves in asymptotically flat space-times,''
Proc. Roy. Soc. Lond. A \textbf{270} (1962), 103-126
doi:10.1098/rspa.1962.0206

\bibitem{Sachs:1962zza}
R.~Sachs,
``Asymptotic symmetries in gravitational theory,''
Phys. Rev. \textbf{128} (1962), 2851-2864
doi:10.1103/PhysRev.128.2851

\bibitem{Strominger:2017zoo}
A.~Strominger,
``Lectures on the Infrared Structure of Gravity and Gauge Theory,''
[arXiv:1703.05448 [hep-th]].

\bibitem{Freidel:2015gpa}
L.~Freidel and A.~Perez,
``Quantum gravity at the corner,''
Universe \textbf{4} (2018) no.10, 107
doi:10.3390/universe4100107
[arXiv:1507.02573 [gr-qc]].

\bibitem{Donnelly:2016auv}
W.~Donnelly and L.~Freidel,
``Local subsystems in gauge theory and gravity,''
JHEP \textbf{09} (2016), 102
doi:10.1007/JHEP09(2016)102
[arXiv:1601.04744 [hep-th]].

\bibitem{Donnelly:2020xgu}
W.~Donnelly, L.~Freidel, S.~F.~Moosavian and A.~J.~Speranza,
``Gravitational Edge Modes, Coadjoint Orbits, and Hydrodynamics,''
[arXiv:2012.10367 [hep-th]].

\bibitem{Freidel:2020xyx}
L.~Freidel, M.~Geiller and D.~Pranzetti,
``Edge modes of gravity. Part I. Corner potentials and charges,''
JHEP \textbf{11} (2020), 026
[arXiv:2006.12527 [hep-th]].

\bibitem{Freidel:2020svx}
L.~Freidel, M.~Geiller and D.~Pranzetti,
``Edge modes of gravity. Part II. Corner metric and Lorentz charges,''
JHEP \textbf{11} (2020), 027
[arXiv:2007.03563 [hep-th]].

\bibitem{Freidel:2020ayo}
L.~Freidel, M.~Geiller and D.~Pranzetti,
``Edge modes of gravity. Part III. Corner simplicity constraints,''
JHEP \textbf{01} (2021), 100
[arXiv:2007.12635 [hep-th]].

\bibitem{DePaoli:2018erh}
E.~De Paoli and S.~Speziale,
``A gauge-invariant symplectic potential for tetrad general relativity,''
JHEP \textbf{07} (2018), 040
[arXiv:1804.09685 [gr-qc]].

\bibitem{Oliveri:2019gvm}
R.~Oliveri and S.~Speziale,
``Boundary effects in General Relativity with tetrad variables,''
Gen. Rel. Grav. \textbf{52} (2020) no.8, 83
[arXiv:1912.01016 [gr-qc]].

\bibitem{Smolin:2003qu}
L.~Smolin and A.~Starodubtsev,
``General relativity with a topological phase: An Action principle,''
[arXiv:hep-th/0311163 [hep-th]].

\bibitem{Freidel:2005ak}
L.~Freidel and A.~Starodubtsev,
``Quantum gravity in terms of topological observables,''
[arXiv:hep-th/0501191 [hep-th]].

\bibitem{MacDowell:1977jt}
S.~W.~MacDowell and F.~Mansouri,
``Unified Geometric Theory of Gravity and Supergravity,''
Phys. Rev. Lett. \textbf{38} (1977), 739
[erratum: Phys. Rev. Lett. \textbf{38} (1977), 1376]

\bibitem{Plebanski:1977zz}
J.~F.~Plebanski,
``On the separation of Einsteinian substructures,''
J. Math. Phys. \textbf{18} (1977), 2511-2520

\bibitem{Aros:1999id}
R.~Aros, M.~Contreras, R.~Olea, R.~Troncoso and J.~Zanelli,
``Conserved charges for gravity with locally AdS asymptotics,''
Phys. Rev. Lett. \textbf{84}, 1647-1650 (2000)
[arXiv:gr-qc/9909015 [gr-qc]].

\bibitem{Aros:1999kt}
R.~Aros, M.~Contreras, R.~Olea, R.~Troncoso and J.~Zanelli,
``Conserved charges for even dimensional asymptotically AdS gravity theories,''
Phys. Rev. D \textbf{62}, 044002 (2000)
[arXiv:hep-th/9912045 [hep-th]].

\bibitem{Gibbons:2004ai}
G.~W.~Gibbons, M.~J.~Perry and C.~N.~Pope,
``The First law of thermodynamics for Kerr-anti-de Sitter black holes,''
Class. Quant. Grav. \textbf{22}, 1503-1526 (2005)
doi:10.1088/0264-9381/22/9/002
[arXiv:hep-th/0408217 [hep-th]].

\bibitem{Frodden:2017qwh}
E.~Frodden and D.~Hidalgo,
``Surface Charges for Gravity and Electromagnetism in the First Order Formalism,''
Class. Quant. Grav. \textbf{35} (2018) no.3, 035002
doi:10.1088/1361-6382/aa9ba5
[arXiv:1703.10120 [gr-qc]].

\bibitem{Frodden:2019ylc}
E.~Frodden and D.~Hidalgo,
``Surface Charges Toolkit for Gravity,''
Int. J. Mod. Phys. D \textbf{29}, no.06, 2050040 (2020)
doi:10.1142/S0218271820500406
[arXiv:1911.07264 [hep-th]].

\bibitem{Durka:2011yv}
R.~Durka and J.~Kowalski-Glikman,
``Gravity as a constrained BF theory: Noether charges and Immirzi parameter,''
Phys. Rev. D \textbf{83} (2011), 124011
doi:10.1103/PhysRevD.83.124011
[arXiv:1103.2971 [gr-qc]].

\bibitem{Durka:2012wd}
R.~Durka,
``Deformed BF theory as a theory of gravity and supergravity,''
[arXiv:1208.5185 [gr-qc]].

\bibitem{Corichi:2016zac}
A.~Corichi, I.~Rubalcava-Garc\'\i{}a and T.~Vuka\v{s}inac,
``Actions, topological terms and boundaries in first-order gravity: A review,''
Int. J. Mod. Phys. D \textbf{25} (2016) no.04, 1630011
doi:10.1142/S0218271816300111
[arXiv:1604.07764 [gr-qc]].

\bibitem{Godazgar:2020gqd}
H.~Godazgar, M.~Godazgar and M.~J.~Perry,
``Asymptotic gravitational charges,''
Phys. Rev. Lett. \textbf{125} (2020) no.10, 101301
doi:10.1103/PhysRevLett.125.101301
[arXiv:2007.01257 [hep-th]].

\bibitem{Godazgar:2020kqd}
H.~Godazgar, M.~Godazgar and M.~J.~Perry,
``Hamiltonian derivation of dual gravitational charges,''
JHEP \textbf{09} (2020), 084
doi:10.1007/JHEP09(2020)084
[arXiv:2007.07144 [hep-th]].

\bibitem{Kijowski:1976ze}
J.~Kijowski and W.~Szczyrba,
``A Canonical Structure for Classical Field Theories,''
Commun. Math. Phys. \textbf{46} (1976), 183-206
doi:10.1007/BF01608496

\bibitem{Crnkovic:1986ex}
C.~Crnkovic and E.~Witten,
``Covariant description of canonical formalism in geometrical theories,'' in {\em Three hundred years of gravitation}, Cambridge University Press, Cambridge U.K. (1989),
Print-86-1309 (PRINCETON).

\bibitem{Crnkovic:1987tz}
C.~Crnkovic,
``Symplectic Geometry of the Covariant Phase Space, Superstrings and Superspace,''
Class. Quant. Grav. \textbf{5} (1988), 1557-1575

\bibitem{Lee:1990nz}
J.~Lee and R.~M.~Wald,
``Local symmetries and constraints,''
J. Math. Phys. \textbf{31} (1990), 725-743
doi:10.1063/1.528801

\bibitem{Iyer:1994ys}
V.~Iyer and R.~M.~Wald,
``Some properties of Noether charge and a proposal for dynamical black hole entropy,''
Phys. Rev. D \textbf{50} (1994), 846-864
doi:10.1103/PhysRevD.50.846
[arXiv:gr-qc/9403028 [gr-qc]].

\bibitem{Wald:1999wa}
R.~M.~Wald and A.~Zoupas,
``A General definition of 'conserved quantities' in general relativity and other theories of gravity,''
Phys. Rev. D \textbf{61} (2000), 084027
doi:10.1103/PhysRevD.61.084027
[arXiv:gr-qc/9911095 [gr-qc]].

\bibitem{Julia:2002df}
B.~Julia and S.~Silva,
``On covariant phase space methods,''
[arXiv:hep-th/0205072 [hep-th]].

\bibitem{Harlow:2019yfa}
D.~Harlow and J.~Q.~Wu,
``Covariant phase space with boundaries,''
JHEP \textbf{10} (2020), 146
doi:10.1007/JHEP10(2020)146
[arXiv:1906.08616 [hep-th]].

\bibitem{GarciaCompean:1999kj}
H.~Garcia-Compean, O.~Obregon, C.~Ramirez and M.~Sabido,
``Remarks on (2+1) selfdual Chern-Simons gravity,''
Phys. Rev. D \textbf{61} (2000), 085022
doi:10.1103/PhysRevD.61.085022
[arXiv:hep-th/9906154 [hep-th]].

\bibitem{Kolanowski:2021hwo}
M.~Kolanowski and J.~Lewandowski,
``Hamiltonian charges in the asymptotically de Sitter spacetimes,''
[arXiv:2103.14674 [gr-qc]].

\bibitem{Misner:1963fr}
C.~W.~Misner,
``The Flatter regions of Newman, Unti and Tamburino's generalized Schwarzschild space,''
J. Math. Phys. \textbf{4} (1963), 924-938
doi:10.1063/1.1704019

\bibitem{Durka:2019ajz}
R.~Durka,
``The first law of black hole thermodynamics for Taub-NUT spacetime,''
[arXiv:1908.04238 [gr-qc]].



\end{thebibliography}
\end{document}